\documentclass[12pt]{iopart}

\usepackage{graphicx}
\newcommand\Rey{\mbox{\textit{Re}}}  % Reynolds number
\renewcommand\Pr{\mbox{\textit{Pr}}}   % Prandtl number
\newcommand\Fr{\mbox{\textit{Fr}}}   % Froude number
\newcommand\Rn{\mbox{\textit{Rn}}}   % 

\def\beg{\begin{eqnarray}}
\def\ende{\end{eqnarray}}

\newsavebox{\astrutbox}
\sbox{\astrutbox}{\rule[-5pt]{0pt}{20pt}}

\newcommand\p{\ensuremath{\partial}}

\def \p   {\partial}

\def \Om  {{\it \Omega}}
\def \Omin  {{\it \Omega}_{\rm in}}
\def \Omout  {{\it \Omega}_{\rm out}}

\def\BV{Brunt-V\"ais\"al\"a~}
\def\curl{\mathop{\rm curl}\nolimits}
\def\div{\mathop{\rm div}\nolimits} 
\def\gsim{\lower.4ex\hbox{$\;\buildrel >\over{\scriptstyle\sim}\;$}} %$
\def\lsim{\lower.4ex\hbox{$\;\buildrel <\over{\scriptstyle\sim}\;$}} %$
\def\rin{\eta}
\renewcommand{\vec}[1]{\mbox{\boldmath $#1$}}

\def\aa{ Astronomy \& Astrophysics\ }
\def\aap{ Astronomy \& Astrophysics\ }

\def\ara\&a{ Ann. Rev. Astronomy Astrophysics}

\def\apj{ The Astrophysical Journal\ }

\def\mnras{Month. Not. Roy. Astr. Soc.\ }

\def\zap{Zeitschrift f. Astrophysik\ }

\begin{document}

\title[Stratorotational instability  in Taylor-Couette flows]{The stratorotational instability of Taylor-Couette flows with moderate Reynolds numbers}

\author{G. R\"udiger$^1$,  T. Seelig$^2$,  M. Schultz$^1$, M. Gellert$^1$, Ch. Egbers$^2$, U. Harlander$^2$},
\address{$^1$Leibniz-Institut f\"ur Astrophysik Potsdam, An der Sternwarte 16, D-14467 Potsdam, Germany}
\address{$^2$Aerodynamik und Str\"omungslehre, BTU Cottbus-Senftenberg, Siemens-Halske-Ring 14, D-03046 Cottbus, Germany}
\ead{uwe.harlander@b-tu.de}

\begin{abstract}
In view of new experimental data the instability against adiabatic nonaxisymmetric  perturbations of a Taylor-Couette flow with an axial density stratification is considered in dependence of
the Reynolds number ($\Rey$) of rotation and the \BV number ($Rn$) of the stratification. The
 flows  at and beyond the Rayleigh limit  become  unstable  between a lower and an upper 
Reynolds number (for fixed $Rn$). The rotation can thus  be too slow or too fast for the stratorotational instability. The upper Reynolds number above which the instability  decays, has its maximum value for the potential flow (driven by cylinders rotating according to the Rayleigh limit) and 
decreases strongly for  flatter rotation profiles finally leaving only isolated islands of instability in the ($Rn/\Rey$) map.  
The maximal possible rotation ratio $\mu_{\rm max}$ only slightly exceeds the shear value of the quasi-uniform flow with $U_\phi\simeq$~const. 

Along and between the lines of neutral stability the wave numbers of the instability patterns for all rotation laws beyond the Rayleigh limit 
 are mainly determined by the Froude number $\Fr$ which is defined by the ratio between $\Rey$ and $Rn$. 
The cells are  highly prolate for $\Fr>1$ so that measurements for too high Reynolds 
numbers become difficult  for axially bounded containers. The instability patterns migrate azimuthally slightly faster than the outer cylinder rotates.
%New  measurements from the SRI experiment at the BTU Cottbus-Senftenberg accurately confirm the predictions of the linear theory. 
\end{abstract}

%\maketitle

%%%%%%%%%%%%%%%%%%%%%%%%%%%%%%%%%%%%%%%%%%
\section{Introduction}
%%%%%%%%%%%%%%%%%%%%%%%%%%%%%%%%%%%%%%%%%%
The magnetorotational instability is now commonly invoked in order to understand the origin of the turbulence in accretion disks.
This is also because  no purely hydrodynamic linear instability exists in accretion disks rotating with the Kepler law  $\Om\propto  R^{-3/2}$. A linear instability is possible, however, when an axial   shear $\Om=\Om(z)$ is included.  In accordance with the stability condition
\beg
\frac{1}{R^4}\frac{\partial}{\partial R} (R^4\Om^2) > \frac{k_R}{k_z}\frac{\partial \Om^2}{\partial z}
\label{intro1}
\ende
\cite[here  written in cylindric coordinates ($R,\phi,z$) and  in terms of the Boussinesq approximation]{UB98}   the flow could  be  unstable for axisymmetric  perturbations with suitable wave numbers $\vec{k}$.  For weak vertical shear the wave numbers  of the possible instability pattern 
must fulfill the condition $k_R\gg k_z$ resulting in cells strongly aligned to the rotation axis.  The  growth time   of this vertical-shear instability, however,  proves to be  longer than the rotation time 
by more than one order of magnitude    \cite{AU04}. The vertical-shear instability which is  closely related to the  Goldreich-Schubert-Fricke  instability \cite{GS67,F68} also has been considered when the axial 
shear is accompanied by an axial density stratification \cite{NG13,SK14,RN16}  resulting in higher growth rates despite of the stabilizing action of the axial entropy gradient. It is obvious that the thermal behavior of the material will play an important role as adiabatic perturbations are much stronger suppressed by the restoring  buoyancy than those with an effective heat transport \cite{U03}.
 
Such purely hydrodynamic instabilities are of particular  interest for cool protostellar disks
where the electric conductivity is  so low that magnetic effects should be  unimportant.  Another
possibility for a hydrodynamic instability is the nonlinear shear instability \cite{CZ03,TC08}.  
However,
a final solution of this problem with numerical  simulations remains difficult because of the large Reynolds numbers required.  The more simple question is 
 whether the direct combination of radial shear  plus density stratification  might  be unstable even  when  both the radial shear {\em and}
 the axial stratification  for themselves are stable.  This, however, is hard to imagine  on the basis  of the  Solberg-H\o{}iland 
conditions 
\begin{equation}  
  \frac{1}{R^3}\frac{\partial (R^4\Om^2)}{\partial R}  - \frac{1}{C_{\rm p} \rho}\frac{\partial P}{\partial z} \frac{\partial S}{\partial  z} >0   
\label{53.3a} 
\end{equation}  
and
\begin{equation}  
 \frac{\partial P}{\partial z} \left(\frac{\partial (R^4\Om^2)}{\partial R} \frac{\partial S } 
{\partial z} - \frac{\partial (R^4\Om^2)}{\partial z} \frac{\partial S}{\partial  
R}\right) < 0   
\label{53.3} 
\end{equation}  
necessary for stability of compressible material. Here $P$ is the pressure, $C_{\rm p}$ the heat capacity for constant pressure and $S$  the specific entropy.  Equation (\ref{53.3a}) provides both  the Rayleigh condition for stability,
$\partial (R^4\Om^2)/\partial R>0$, for isentropic axial stratification and the Schwarzschild criterion for stability, $\partial S/\partial z>0$, for resting fluids. If, however,  a fluid   with  $\partial P/\partial z < 0$  rotates  with  a stable rotation law $\Om=\Om(R)$, then also the second stability  Solberg-H\o{}iland condition  (\ref{53.3})  is  fulfilled with 
 the Schwarzschild  criterion 
 %in the traditional form 
\beg 
\frac{\partial S}{ \partial z} > 0
\label{Schwarz}
\ende 
 \cite{LS77,ER89}.  A combination of stable  axial density stratifications and centrifugally-stable rotation laws should thus also be stable. Note, however,  that  Eqs. (\ref{53.3a}) and (\ref{53.3})  are only valid  for stability against axisymmetric perturbations within the short-wave approximation \cite{T78}. 
 
 %Because of  (\ref{Schwarz}) unstable compressible models must not be adiabatic and/or do not fulfill the short-wave conditions -- or the instability is driven by the boundary conditions.

The stabilizing role of  axial density stratification  has indeed been observed in several  
Taylor-Couette experiments 
with stationary outer cylinder  and for  incompressible fluids   \cite{BG95,HL97,CJ00}.
 The axial density stratification increases the critical Reynolds numbers (of the inner cylinder) for   the onset of the  axisymmetric mode. With numerical simulations \cite{HL97}
 found  the first unstable mode indeed  to be axisymmetric if the Prandtl number has been chosen high enough. One can  show that 
 the excitation of this mode needs increasingly higher Reynolds numbers if also the outer cylinder rotates so that it goes to infinity if 
  $\Om \propto 1/R^2$ is approached   \cite{SR05}.
This rotation law is curl-free ($\curl \vec{U}=0$,  called the  `potential' flow) and separates the
  centrifugally unstable flows  from the  stable flows defining  the Rayleigh limit. These findings are in  agreement with the criterion (\ref{Schwarz}).
  However,  under the presence of an axial density
stratification  persisting  { \em nonaxisymmetric disturbances} for the stable flows   slightly  beyond the Rayleigh limit  have been observed \cite{WC74}.  
 Numerical studies also revealed    that the combination of convective-stable axial density stratification  and centrifugally-stable
differential rotation  is linearly unstable against nonaxisymmetric perturbations 
\cite{MM01,YM01,SR05,U06}. Obviously, the criteria (\ref{53.3a}) and (\ref{53.3}) are  only necessary for stability but not sufficient.
 This instability against nonaxisymmetric modes  (now
called  the stratorotational instability, SRI) does not require the presence of radial boundaries \cite{DM05}. 

Almost all previous  calculations concerned Boussinesq fluids (with infinite speed of sound) fulfilling   the incompressibility condition $\div \vec{U}=0$ (see, however \cite{U06}). The density fluctuations are assumed as   due to the action of fluctuating flows within a medium of axially stratified background density. A possible molecular density diffusion $\kappa$ is often ignored in the calculations because of its smallness leading to Schmidt  numbers exceeding  $\mathcal{O}(10^2)$ \cite{HL97,CJ00,LP16}. 
For too small Schmidt   number the diffusion suppresses  density perturbations.
Following \cite{HL97} we shall use in this paper  the notation Prandtl number for the ratio of viscosity $\nu$ and density-diffusivity $\nu/\kappa$ instead of  Schmidt number because of its identical definition if a temperature  stratification is used instead of the density stratification.

The {\em inviscid} equations of  \cite{YM01} for the potential flow combining
a rigid inner boundary with an infinite gap width have been solved  showing that the most unstable modes belong to high azimuthal wave numbers   \cite{LR10}  .  Weak
stratification suppresses the instability. Pure Keplerian rotation  -- which is not an exact solution of the zero-order equations -- proves to be
stable for disturbances with  azimuthal mode number smaller than 14.  The same system of inviscid equations is also reported as allowing instability if the
angular velocity of the rotation increases outwards ('superrotation', see \cite{PB13}).
Taylor-Couette flows with {\em finite viscosity} have been considered  even  formulating     a stability 
criterion  \cite{SR05}.  For a medium gap  and a fixed small Froude number of 0.5 the flattest unstable rotation law 
was $\Om\propto 1/R$ which has experimentally been confirmed by \cite{LL07}.  Later calculations 
demonstrated that for unity Froude number  indeed $\Om\propto 1/R$ approaches the smoothest unstable rotation law for a wide variation of 
the gap widths  \cite{RS09}. 
 \cite{IS16} reported new  experimental results   with a small-gap container  with variable
rotation laws including the existence of minimal and maximal  values  of the Reynolds number between them the SRI exists. We shall here discuss several of their  findings in the light of  new  experimental data and theoretical results also considering  the critical Reynolds number for fixed density stratification as a function of the ratio 
\beg
\mu=\frac{\Omout}{\Omin}
\label{ratioom} 
\ende
of the rotation rates of the outer and  the inner cylinder.

The paper is organized as follows. In Section 2 we briefly describe the experimental setup and in Section 3  details about the Taylor-Couette flow are given
model that is linearized and solved numerically. This is done first in Section 4 where we present stability properties for the most prominent
example of potential flow (`the Rayleigh limit'). Subsequently, in Section 5
we do the same for flat rotation laws and compare stability diagrams and drift rates with the ones obtained from experimental measurements. 
Moreover, in Section 6 we discuss the axial wave number of the unstable modes and give limits for the Froude number in view of the
finite size of the experimental apparatus. In Section 7  the growth rate of the instability is shortly commented. 
%and in Section 8 one finds discussions and conclusions.

%%%%%%%%%%%%%%%%%%%%%%%%%%%%%%%%%%%%%%%%%%%%%%%%%%%%%%%%%%%%%
\section{Experimental realization}\label{experiment}
\begin{figure}
 \centering
 \includegraphics[width=0.33\textwidth]{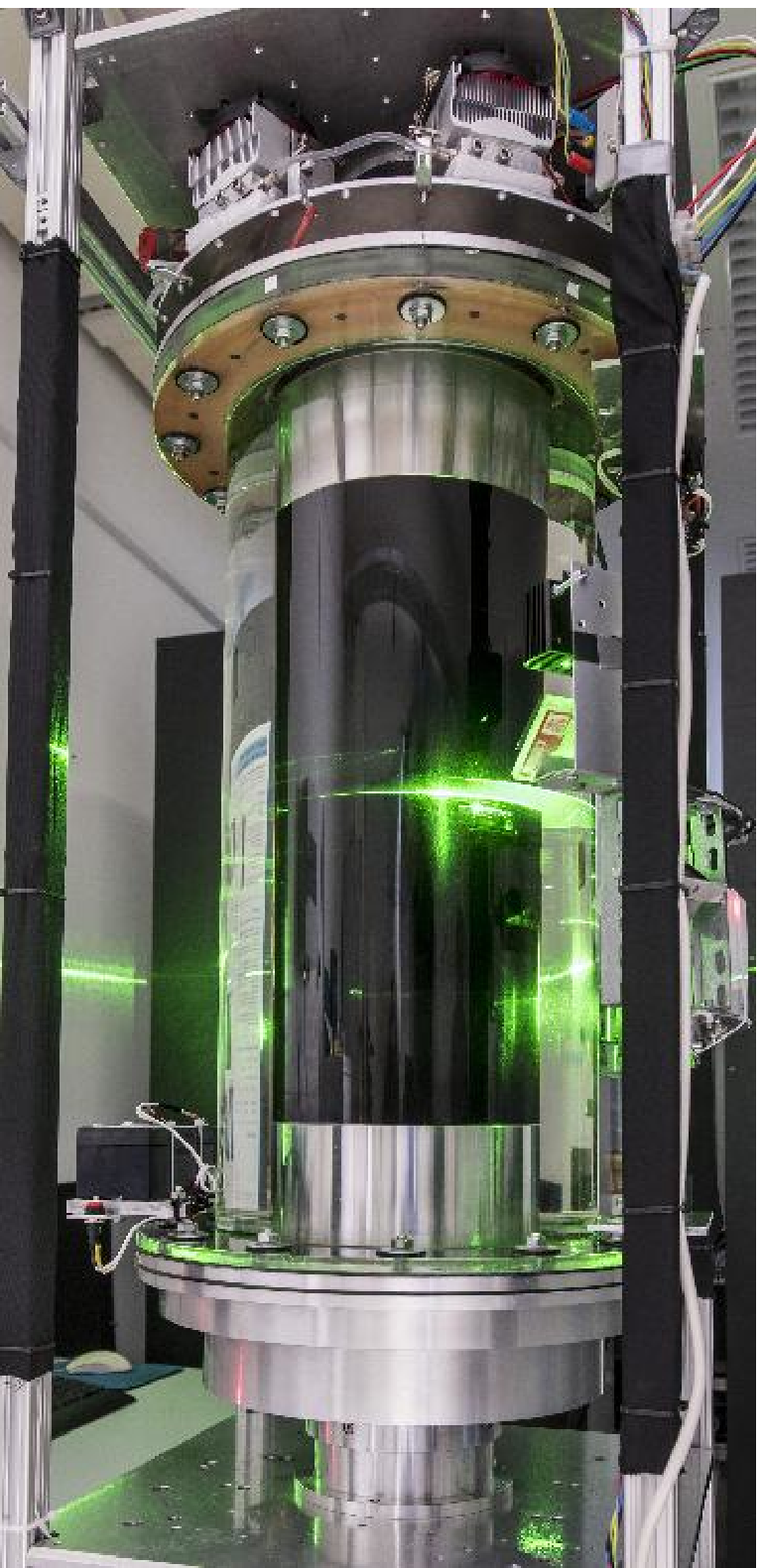}
 \hspace{1.5cm}
 \includegraphics[width=0.43\textwidth]{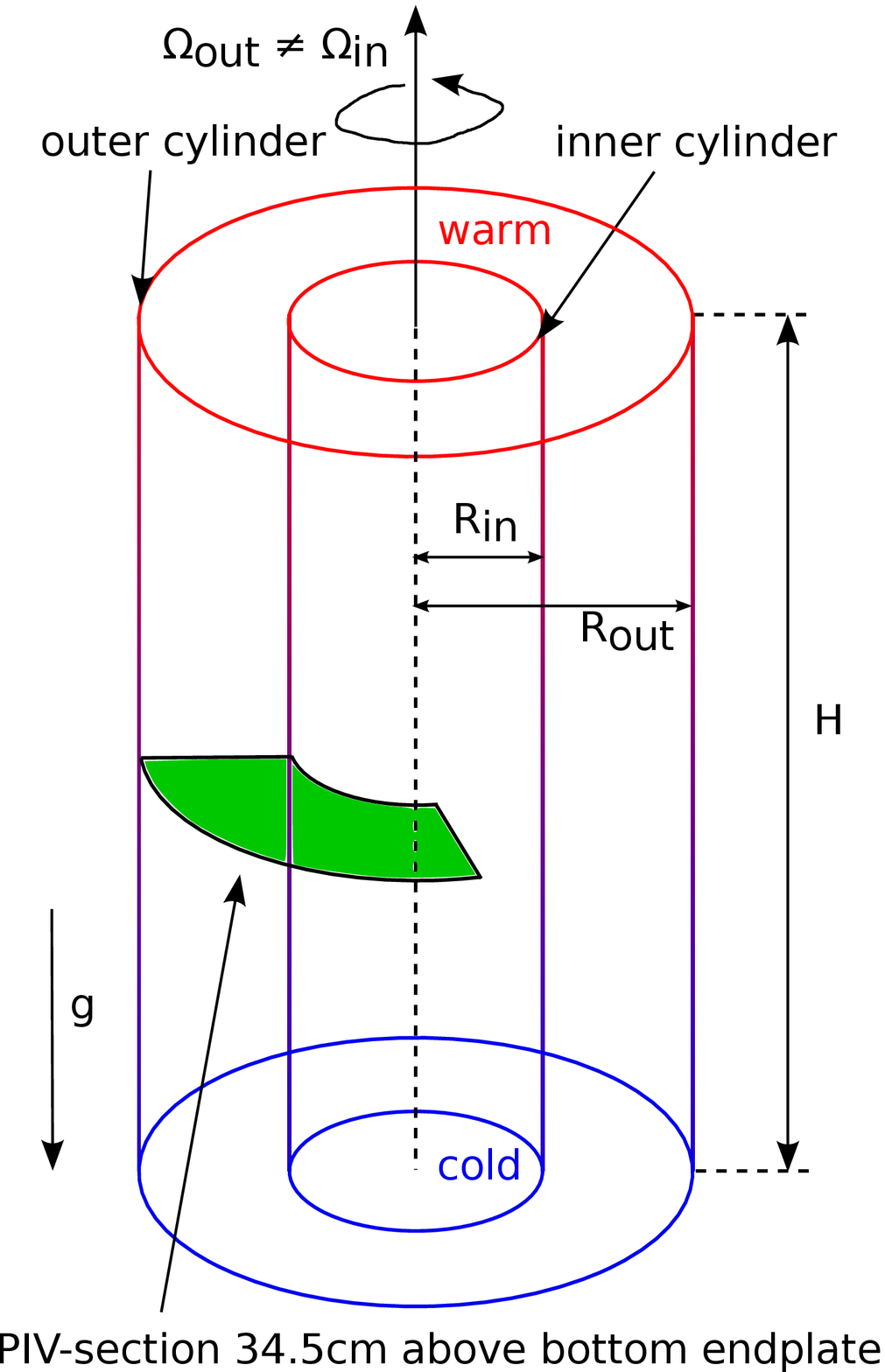}
 \caption{Left: The SRI experiment at the Brandenburgische Technische Universit\"at  Cottbus-Senftenberg. The temperature
difference between top and bottom varies from 1 to 5 K. Right: Sketch of the 
          experimental cell and position of the PIV-section used to determine the flow state  and the drift rate.  It is  $R_{\rm in}=7.5$ cm, $R_{\rm out}=14.5$ cm. $H=70$~cm, $\eta=0.52$.}
 \label{setup}
\end{figure}

%%%%%%%%%%%%%%%%%%%%%%%%%%%%%%%%%%%%%%%%%%%%%%%%%%%%%%%%%%%%%
The SRI experiment of the BTU Cottbus-Senftenberg is used to probe the main results of the presented  stability analysis.  In
opposition to earlier experiments it works with a stable stratification due to a positive axial temperature gradient ${\rm d} T/{\rm d} z$ 
by heating the cylindrical gap from {\em above}, see \cite{GR09}. Instead of the density gradient it is here the temperature 
gradient that defines the \BV frequency. 
Beyond the Rayleigh limit   a
destabilization happens already for a temperature difference between top and bottom of a few K. The container has an aspect ratio of
$\Gamma=H/(R_{\rm out}-R_{\rm in})=10$  and a ratio  
\beg
\eta=\frac{R_{\rm in}}{R_{\rm out}}
\label{ratiora} 
\ende
of the two  cylinder radii $R_{\rm in}$ and  $R_{\rm out}$  of $\eta=0.52$  (Fig. \ref{setup}). By the   rotation 
the instability cells become axially  elongated. We shall show that an  aspect ratio of only  $\Gamma=10$  does not  allow too fast rotation.    The outer cylinder is made
of acrylic glass and accessible for Particle Image Velocimetry (PIV). The inner cylinder is made of anodized aluminum in order to suppress
disturbing optical reflection. Both horizontal endplates are connected to the outer cylinder.  

\begin{figure}
 \centering
 \includegraphics[width=10cm]{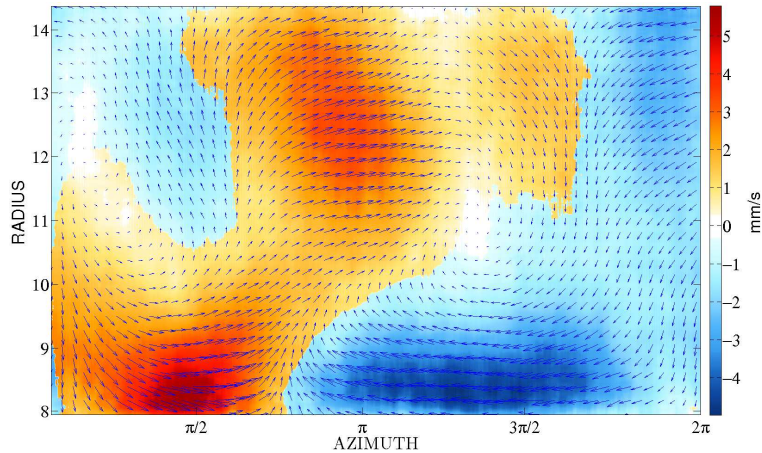}
 \caption{
 %Left:  Power spectrum deduced from PIV  data with several distinct peaks. Right:   
 Timeseries associated with the period/azimuth of mode $m=1$ of the nonaxisymmetric PIV velocity field  in the ($R/\phi$) plane  with dominating $m=1$ mode and one cell in radius. Note the azimuthal  phase shift of the maxima indicating the existence of two vortices shifted in azimuth by 180$^\circ$. A  harmonic reconstruction with  single peak frequencies provides   the energy of the modes  with various $m$. The whole  pattern   drifts  in the direction of increasing azimuth. $\mu=0.45$, $\Rey=406$,  $Rn=206$. }
 \label{pattern}
\end{figure}

On the upper endplate 12 Peltier elements have been installed. Six of them operate as heat source and were mounted in hexagonal order at
the inner part of the plate. All further Peltier elements work as cooling system. They are mounted on top of aluminum blocks. Water is
cooled inside the aluminum blocks and continuously pumped downward into a reservoir mounted below the bottom aluminum endplate to cool the 
endplate and keep its temperature constant.
The inner and outer cylinders are driven independently by two DC motor units.  The axial temperature profile is analyzed with an infrared
camera. Flow velocity measurements are done with a corotating mini-PIV system  with a green laser module and a corotating camera. The tank
is filled with silicone oil M5 with   $\Pr=57$.  
Figure \ref{setup} shows  the position and extension of the PIV section 34.5cm above the cooled bottom endplate.
We calibrated the PIV section with a chess board pattern.
Velocity fields are computed by the use of MatPIV~\cite{SV04}.
Initially, the velocity components are given in Cartesian coordinates.
The origin of the PIV segment, needed for polar coordinate transformations, is determined by applying the calibration grid.
In a next step the grid and velocity components have been transformed into polar coordinates.
On a line of constant angle and varying radial gridpoints ($\phi=0$, central part of the PIV-section) we stored the polar velocity components as time series.
We next compute power spectra for each set of PIV data. Transition from a stable to an unstable flow is signaled by the occurrence of isolated peaks in the spectra. In a subsequent step we reconstruct the SRI modes by applying a least-square harmonic fit to  the PIV time series using the peak frequencies of the power spectra.  In that way we obtain the spatial structure of the modes  (the  azimuthal wave numbers) and their development in time  (the azimuthal drift rates of the patterns).  
Figure  \ref{pattern} shows  a  SRI velocity field $\vec{u}=(u_{R},\, \Delta u_{\phi}=u_{\phi}-\bar{u}_{\phi}^{t}(R) )^{T}$ in the azimuthal-radial plane measured with the  PIV. The magnitude $\mid\vec{u}\mid=(u_{R}^{2}+\Delta u_{\phi}^{2})^{1/2}$ is shown in red if $\Delta u_{\phi}$ is positive (prograde) and in blue if $\Delta u_{\phi}$ is negative (retrograde).
The  pattern will be decomposed into single modes with  different azimuthal wave numbers from which  the azimuthal drift of each single  mode can be derived. More details about the experiment and the data analysis are given by \cite{SB16}.  
 
\begin{figure}
 \centering
 \includegraphics[width=0.65\textwidth]{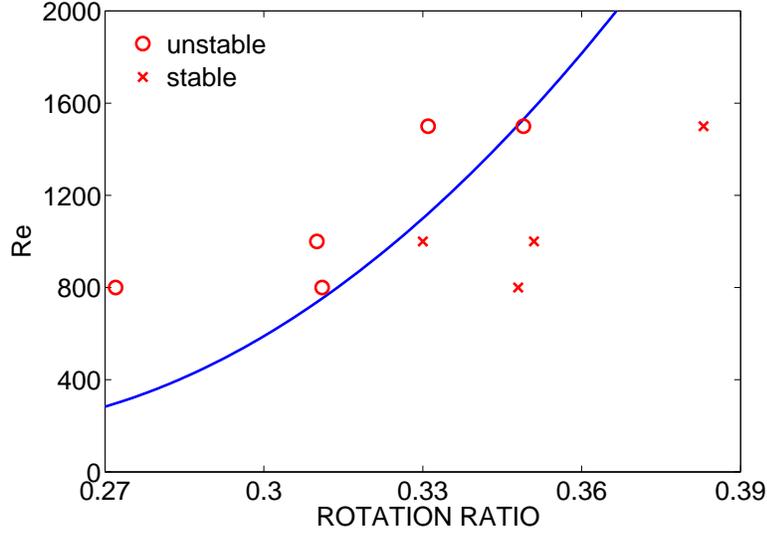}
 \caption{ $\Rey$ versus  $\mu$ for  Taylor-Couette  experiments  without stratification. The endplate effects lead to a nonaxisymmetric   instability pattern (red circles)  due to the shear-layer instability. All  crosses representing stability are located right of the theoretical curve while the circles are located at the curve or left of it. Close to the Rayleigh limit at $\mu=0.27$ this instability appears already for low Reynolds numbers.   Numerical simulations with (!) endplates provide the solid blue line.}
 \label{shear_inst}
\end{figure}

Taylor-Couette  systems with endplates generate  strong shear layers. The resulting meridional circulation modifies the rotation profile close to  the endplates. For strong enough differential rotation this shear layers become unstable
and nonaxisymmetric modes  develop \cite{AG08,A12}. Adding stratification suppresses this instability and reduces the 
shear layer and the related circulations  \cite{LP16}. Both effects might appear 
in the experiment and it is difficult to say how strong the stratification must be in order to stabilize the  shear-layer instability.  We therefore probed the stability of  
{\em unstratified} flows beyond the Rayleigh limit which all should be stable against axisymmetric and nonaxisymmetric perturbations. The results are shown in Fig. \ref{shear_inst} in comparison with nonlinear numerical simulations  by 
\cite{GR09}. The unstable flows in the experiment are found above the line that marks the onset of the shear-layer instability  in simulations. The critical Reynolds numbers for  flows near the Rayleigh limit  are rather low (of order 10$^3$)  so that only experiments with  flatter rotation laws appear to be   reasonable.

%%%%%%%%%%%%%%%%%%%%%%%%%%%%%%%%%%%%%%%%%%%%%%%%%%%%%%%%%%%%%%%%
\section{The Taylor-Couette flow model}\label{sec_model}
%%%%%%%%%%%%%%%%%%%%%%%%%%%%%%%%%%%%%%%%%%%%%%%%%%%%%%%%%%%%%%%%
We have to formulate  the basic state with prescribed  velocity profile $\vec{U}=(0,R\Om,0)$, given rotation law $\Om=\Om(R)$ and density
axial stratification $\rho_0(z)$. The equations for the background density and azimuthal flow profile  are
\beg
{ \frac{U_\phi^2}{R}=\frac{1}{\rho_0}\frac{\partial P}{\partial R} ,
\quad\quad\quad\quad
\frac{1}{\rho_0 }\frac{\partial P}{\partial z} = - g,}\quad\quad\quad\quad
{ \frac{\partial^2 U_\phi}{\partial R^2}+\frac{1}{R}
\frac{\partial U_\phi}{\partial R} - \frac{U_\phi}{R^2} =0.}
\label{sysa}
\ende
The last equation yields 
\begin{equation}
\Om=a+\frac{b}{R^2}
\label{Om} 
\end{equation}
as the solution for the angular velocity where $a$ and $b$ are free constants. The last term in (\ref{Om}) is curl-free  (the 
`potential' flow) defining  the Rayleigh limit in the theory of the Taylor-Couette flows. The two free constants must  be fixed 
by  the boundary values $\Om_{\rm in}$ and $\Om_{\rm out}$ of the angular velocity of the inner cylinder (radius $R_{\rm in}$) and the outer
cylinder (radius $R_{\rm out}$). It follows           
\begin{equation}
a=\Om_{\rm in}\frac{\mu -  \eta^2}{1-\eta^2}, \quad\quad\quad
b=\Om_{\rm in}R_{\rm in}^2\frac{1-\mu}{1-\eta^2}
\label{ab}
\end{equation}
with the ratios (\ref{ratioom}) and (\ref{ratiora}).
%\begin{equation}
%\mu =\frac{\Om_{\rm out}}{\Om_{\rm in}}, \quad\quad\quad
%\eta =\frac{R_{\rm in}}{R_{\rm out}}.
%\end{equation}
The Rayleigh limit  is defined by $a=0$, i.e. $ \mu=\eta^2$. A quasi-Keplerian flow results for $\mu=\eta^{1.5}$ while $\mu=\eta$ leads to a flow with quasi-uniform linear velocity $U_\phi$.

From (\ref{sysa}) follows
\beg
R\Om^2 \frac{\partial \rho_0}{\partial z}
+g\frac{\partial \rho_0}{\partial R} =0
\label{cond1}
\ende
if the rotation law is forced to be independent of $z$. One finds 
$\partial \rho_0/\partial R = - \epsilon \partial \rho_0/\partial z$ with the ratio $\epsilon = R\Om^2/g$ of centrifugal acceleration to 
the gravity. Only for small $\epsilon$ in rotating fluids  the radius-dependence of $\rho_0$ can be neglected. 

The Froude number $\Fr=\Om_{\rm in}/N$ as the ratio between the rotation rate $\Om_{\rm in}$ of the inner cylinder and  the \BV frequency   
\begin{equation}
 N = \sqrt{- g \frac{\p \log \rho_0}{\p z}}
\end{equation}
describes a normalized rotation rate. The  Reynolds  number $\Rey$ and the \BV number $\Rn$ are  defined by 
\begin{equation}
\Rey=\frac{\Om_{\rm in}R_{\rm in} D}{\nu}, \quad\quad\quad\quad \Rn=\frac{N R_{\rm in} D}{\nu},
\end{equation}
%and the \BV number is 
%\begin{equation}
%\Rn=\frac{N R_{\rm in} D}{\nu},
%\end{equation}
hence $\Fr=\Rey/\Rn$. The gap width is $D=R_{\rm out}-R_{\rm in}$.

For the  normalized centrifugal acceleration $\epsilon$  one finds
$
\epsilon=\simeq   \Fr^2 {R g}/{c^2_{\rm ac}}
$
with $c_{\rm ac}$ as the speed of sound which for water is more than 10$^5$ cm/s. Hence, $Rg/c^2_{\rm ac}$ is  $\mathcal{O}(10^{-6})$ for $R\simeq 10$\ cm. 
The condition $\epsilon\ll 1$ is thus always fulfilled so that the centrifugal acceleration can be neglected in Eq. (\ref{cond1}).

Consider the basic flow, density  and  pressure as perturbed to a flow $\vec{U}+\vec{u}$, density $\rho_0+\rho$ and $P+p$. In cylindric geometry the Boussinesq form of the system of hydrodynamic equations  for the perturbations $\vec{u}$ and $p$ is
\beg
{ \frac{\partial u_R}{\partial t} + \Om \frac{\partial u_R}{\partial
\phi}- 2\Om u_\phi=
-\frac{\partial }{\partial R}\left(\frac{p}{\rho_0}\right)}
 + \nu \left( \Delta u_R-\frac{2}{R^2}\frac{\partial u_\phi}{\partial \phi}-
\frac{u_R}{R^2} \right),\nonumber\\
{ \frac{\partial u_\phi}{\partial t} + \Om \frac{\partial u_\phi}{\partial
\phi}+\frac{1}{R}\frac{\partial R^2 \Om}{\partial R}u_R=
-\frac{1}{R}\frac{\partial }{\partial \phi}\left(\frac{p}{\rho_0}\right)+}
 \nu \left( \Delta u_\phi+\frac{2}{R^2}\frac{\partial u_R}{\partial \phi}-
\frac{u_\phi}{R^2} \right),\nonumber\\
{ \frac{\partial u_z}{\partial t} + \Om \frac{\partial u_z}{\partial
\phi}=-\frac{\partial }{\partial z}\left(\frac{p}{\rho_0}\right)
-g\frac{\rho}{\rho_0} + \nu \Delta u_z,  }\ \ \ \ \ \ \ \ \ \ \ \ \ \ \ \ \ \ \ \  \nonumber
 \\
 \frac{\partial }{\partial t}\left(\frac{\rho}{\rho_0}\right)
+\Om \frac{\partial }{\partial \phi}\left(\frac{\rho}{\rho_0}\right)
-\frac{N^2}{g}u_z=0 \ \ \ \ \ \ \ \ \ \ \ \ \ \ \ \ \ \ \ \ \ \ \ \ \ \ \ \ 
\label{sysbo}
\ende
with $
\Delta F={\partial^2 F}/{\partial R^2}+({1}/{R})
{\partial F}/{\partial R}+({1}/{R^2}){\partial^2 F}/{\partial
\phi^2}+{\partial^2 F}/{\partial z^2}
$ and for incompressible fluids, $\div{\vec{u}}=0$. The $\rho_0$ in the system denotes a constant reference value of the density. The last equation in (\ref{sysbo}) describes the generation of density fluctuations by a flow field in combination with  an axial gradient of the background density. The adiabatic temperature equation used in the presence of a prescribed temperature gradient has the identical form, see \cite{GR09}.

As we  shall always consider the \BV frequency $N$ as uniform the coefficients of the system only depend on the radial coordinate, so that a normal mode expansion of the
solutions, 
\beg
F(R,\phi,z,t)=F(R){\textrm{e}}^{{\rm i}(m\phi+kz-\omega t)},
\label{Fourier}
\ende
can be applied with the azimuthal wave number $m$, the axial wave number $k$ and the complex Fourier frequency $\omega$ for all fluctuating quantities. The no-slip boundary
conditions at the inner and outer cylinder ($u_R=u_\phi=u_z=0$) complete the  eigenvalue problem. Note that there is no molecular density diffusion  $\kappa$  in the system hence the microscopic Prandtl number  ${Pr}=\nu/\kappa$ is assumed as infinity. \cite{PB13} work with a ratio of the diffusivities of $700$ which certainly  allows the  adiabatic approximation (\ref{sysbo}). The Prandtl number of the silicon oil used in the apparatus shown in Fig. \ref{setup} is 57. The condition that in the last relation of (\ref{sysbo}) a molecular diffusion term  can be neglected against the drift term of the nonaxisymmetric instability pattern simply reads $m \Rey>1/{Pr}$ which is always fulfilled  for Prandtl  numbers exceeding unity. 

A linear code is used to solve the set of ordinary differential equations for the radial profiles of flow, density and pressure.  The solutions are optimized with respect to that axial wave number $k$ which
provides the lowest Reynolds number. The wave numbers are normalized with the characteristic radius $R_0=\sqrt{R_{\rm in}D}$ (which
for $\eta=0.52$ is very close to the gap width $D$) and the   frequencies $\omega$ with the rotation rate  $\Om_{\rm in}$ of the inner
cylinder. As described by \cite{SR05}  the code has been tested with  data of  \cite{WC74}.

We shall start with the discussion of the conditions for neutral stability in the ($\Rn/\Rey$) plane of the  modes. Neutral
instability implies  vanishing growth rates. The shear number $\mu$ is considered as the free parameter.

%%%%%%%%%%%%%%%%%%%%%%%%%%%%%%%%%%%
\section{Potential flow}
\begin{figure}
\centering
\includegraphics[width=0.85\textwidth]{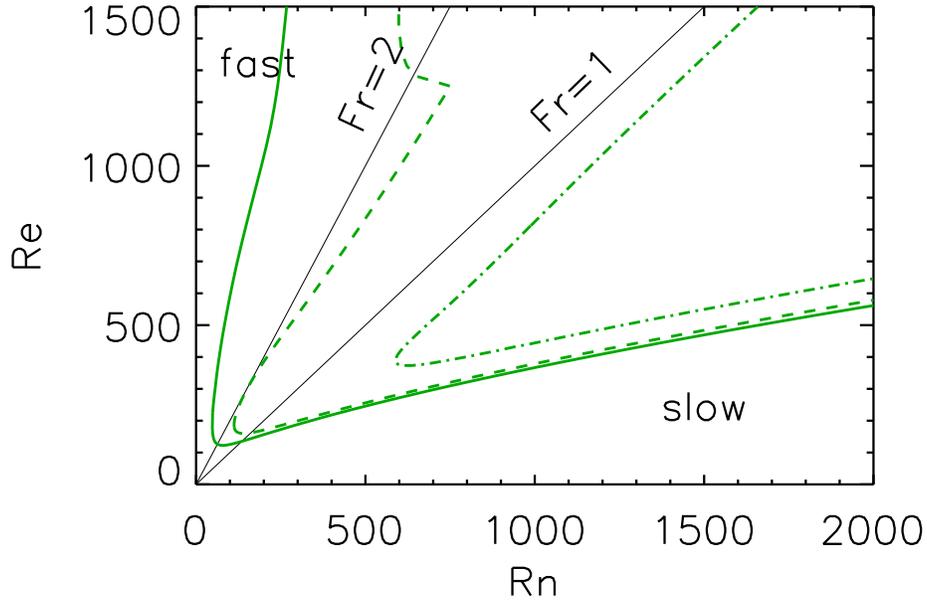}
\caption{Stability map for the modes with $m=1$ (solid), $m=2$ (dashed) and $m=3$ (dot-dashed) in the ($\Rn/\Rey$) plane for the potential flow. The 
         gray solid lines display $\Fr=1$ and $\Fr=2$. $\eta=0.52$,  $\mu=0.27$.}
\label{fig1}
\end{figure}

%%%%%%%%%%%%%%%%%%%%%%%%%%%%%%%%%%%
 The most prominent example for
the instability is the potential flow with $\mu=0.27$. \cite{LR10} argue that the instability of the stratified potential flow is the
most unstable one in agreement with our numerical results.  
%The models  also suggest that all rotation laws with $\mu\lsim 0.42$ (the quasi-Keplerian flow with $\mu=0.37$ included) belong to this instability class unless  the Reynolds numbers lie below its lower limit or above its upper limit..
Figure \ref{fig1}  shows 
the lines of marginal instability derived from    (\ref{sysbo}) for  vanishing imaginary part of $\omega$.  The potential flow proves to be  unstable against
nonaxisymmetric perturbations with low mode numbers $m$. The lines of neutral stability in the given  sector   are almost straight lines 
%which, however, cannot simply  be described by  isolines of  $\Fr$.
so that  a  { necessary} condition  $\Fr< \Fr_{\rm max}$ for instability of the potential flow  against nonaxisymmetric perturbations is suggested with   $\Fr_{\rm max}\simeq 5.5$ for $\Rey\lsim 1000$.  The 
  values for the non-potential flows with weaker shear, however, will be basically smaller. Figure  \ref{fig3}  shows that for $\Rey>1000$   the values  of $\Fr_{\rm max}$ start to depend on the Reynolds number. The stability curves in the full ($\Rn/\Rey$) planes  are   more complex to be described  in terms of a single $\Fr$.  
 
The main result from  Fig. \ref{fig1} is that the instability domain forms a cone   in the ($\Rn/\Rey$) plane with an upper fast-rotation bound and a lower slow-rotation bound.  The stability curves always possess a minimum $\Rn$. For larger  $\Rn$ there are always two Reynolds numbers $\Rey$ limiting instability cone. The rotation can thus be too slow  and  it  can be too fast for SRI. The difference between the maximum
and the minimum $\Rey$ increases for increasing $\Rn$.  For $\Rey$  larger than its absolute minimum value the line $\Fr=1$ always belongs to the instability domain.  Also no crossing points of the two lines of neutral instability have been found for the potential flow with the straight  lines $\Fr=1$ or $\Fr=2$. The form of the instability curves suggests   that this condition also holds for very large Reynolds numbers, i.e.
in the diffusion-less limit $\nu\to 0$. We shall see below that this statement does not remain true if more flat rotation laws are considered.
 
Figure \ref{fig1}  also shows the instability maps of the modes with $m=2$ and $m=3$  which both are  close to the curve for $m=1$. The
latter encloses the curve for $m=2$ which itself encloses the curve for $m=3$. For the upper branch in the  ($\Rn/\Rey$) plane
representing the solutions for fast rotation the different   lines are easy to see while they are much more  close together at the lower branch. 
This means that strong shear has a  stronger stabilizing effect to the modes with higher $m$ than for $m=1$. We shall  see below that this mechanism is even more effective for flatter rotation laws.
 \begin{figure}
\centering
%\hbox{\includegraphics[width=0.50\textwidth]{fig3a.eps}
\includegraphics[width=0.49\textwidth]{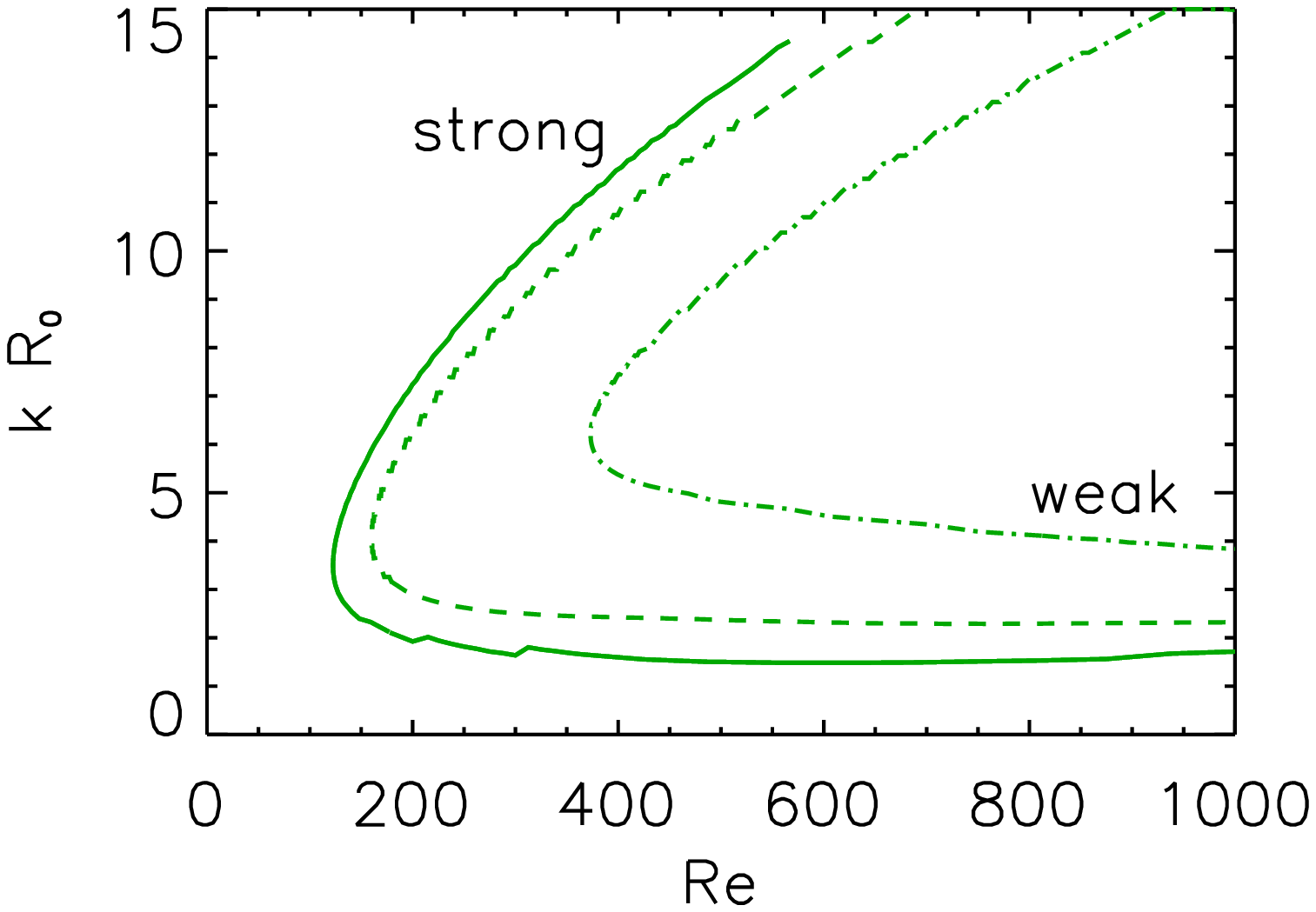}
\includegraphics[width=0.49\textwidth]{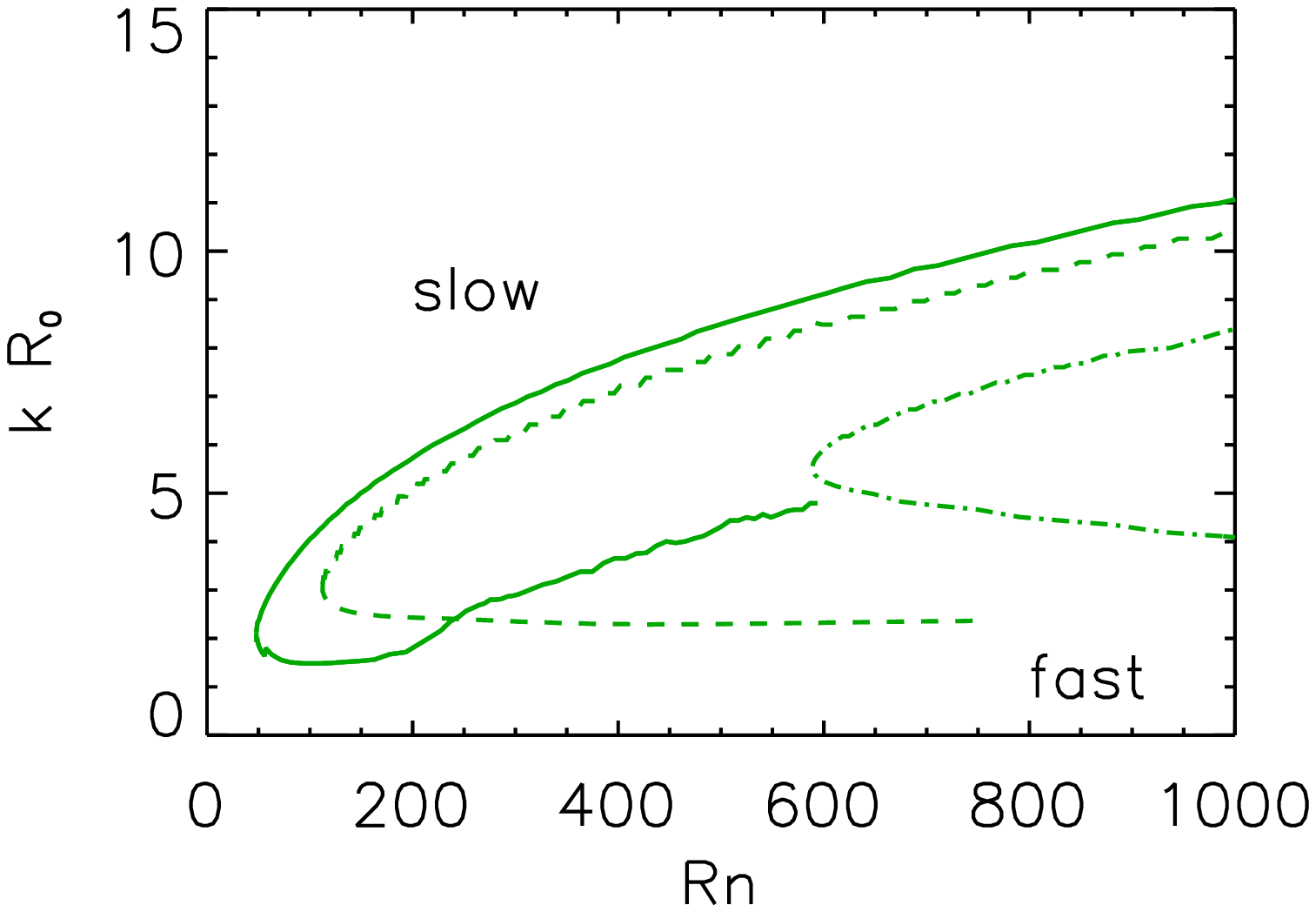}
\caption{Axial wave numbers $k R_0$ along the lines of neutral instability shown in  Fig. \ref{fig1} for $m=1,2,3$ as functions of $\Rey$ (left) and $Rn$ (right). The upper branches of the lines  belong to 
          strong stratification (left) and/or to slow  rotation (right). Strong stratification and/or  slow rotation  lead to high wave numbers and vice versa. $\eta=0.52$,  $\mu=0.27$.}
\label{fig1b}
\end{figure} 

In  Fig. \ref{fig1b} the normalized axial  wave numbers along the lines of neutral stability are given for the modes $m=1,2,3$ as a function of the Reynolds number (left) and of the \BV number (right). The lower
parts of the curves belong to the upper branches  in the stability map  while the upper parts in the two panels represent the lower branches in the instability map.  
For fixed Reynolds number weak stratification leads to small axial wave numbers and strong stratification leads to large wave numbers. The action of the Taylor-Proudman theorem 
can be recognized  by the right panel of Fig. \ref{fig1b}.  For fixed stratification fast rotation produces small axial wave numbers and slow rotation produces large wave numbers. 
%Along the fast rotation branch, therefore, the cells are aligned with the rotation axis. 
Note that at the beginning of the  fast-rotation branch of the potential flow in Fig. \ref{fig1b}  the wave numbers are  almost constant with $k R_0\simeq 2$.

\begin{figure}
\centering
\hbox{\includegraphics[width=0.5\textwidth]{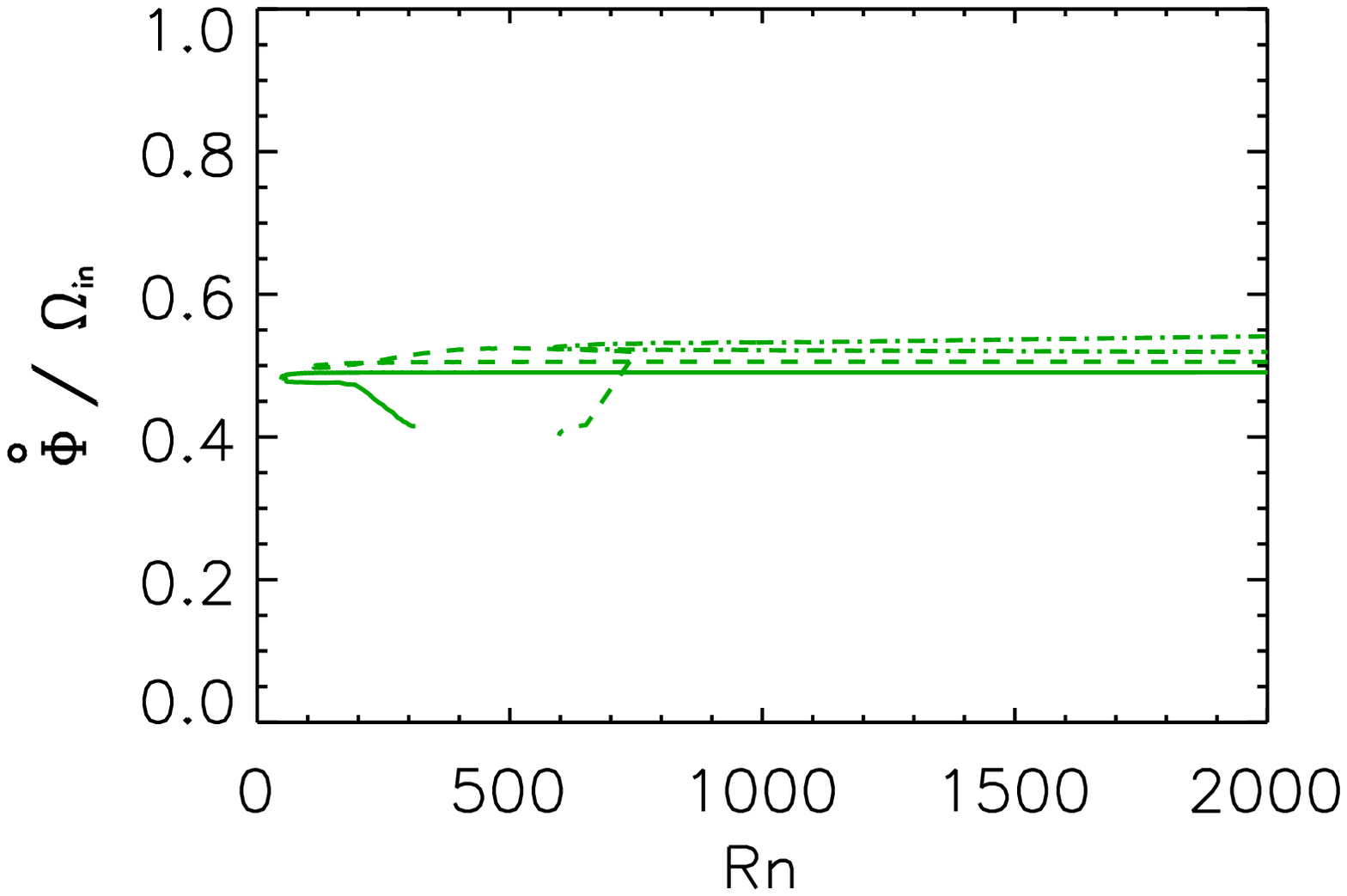}
      \includegraphics[width=0.51\textwidth]{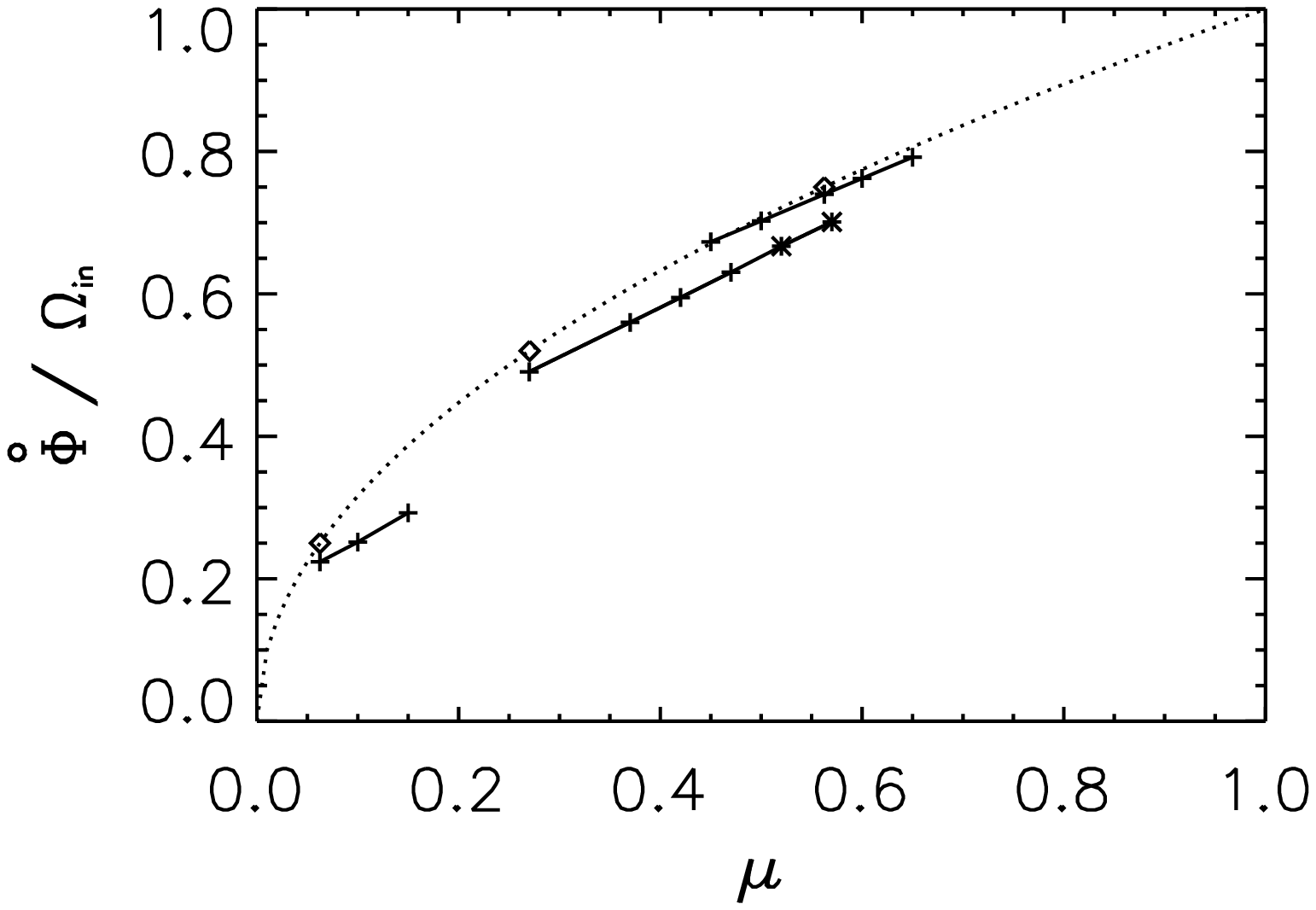}
     }
\caption{Left: Drift values (\ref{drift})  of the nonaxisymmetric modes with  $m=1,2,3$ (solid/dashed/dot-dashed)    for the potential flow with   $\eta=0.52$ and $\mu=0.27$. The data belong to the slow-rotation branches and to  the fast-rotation branches in Fig. \ref{fig1}.  
         Right: Drift values  for $m=1$ as function of the rotation ratio  $\mu$ for three different gaps  with $\eta=0.25$, $\eta=0.52$,  $\eta=0.75$ (from left to 
         right). The dotted curve gives   $\dot\phi/\Om_{\rm in}=\sqrt{\mu}$ for the potential flow of the three models. }
\label{fig2}
\end{figure}

The equation system (\ref{sysbo}) only possesses solutions for finite values of the Fourier frequency $\omega$. Its imaginary part provides the growth or decay rate of the solution while  its real part describes an azimuthal drift of the entire  nonaxisymmetric instability  pattern. It is here  given   by the real  part
$\omega_{\rm dr}$ of the frequency $\omega$ of the Fourier expansion (\ref{Fourier}) normalized with the rotation rate of the  inner cylinder. Because of 
\beg
 \frac{ \dot \phi}{\Om_{\rm in}}=\frac{\omega_{\rm dr}}{m \Om_{\rm in}}
 \label{drift} 
\ende
the azimuthal drift has the same sign of  $\omega_{\rm dr}$. As the drift rates  in Fig. \ref{fig2} (left panel)
are always positive the patterns  always migrate in positive $\phi$-direction. Note that   a drift value of $\mu$   describes  exact corotation of the instability  pattern with the {\rm outer} cylinder. The positive  deviation  from this  value   makes
the pattern  slightly  faster rotating than the outer cylinder. 

The drift (\ref{drift}) for the potential flow (along the slow-rotation branch of the instability cone) almost  proves  to be independent of $\Rn$. The
numerical values only  show a  slight increase towards large $\Rn$ (Fig. \ref{fig2}, left). %It is thus suggestive here  that $\lim \dot \phi/\Om_{\rm in}= \eta$ for
%${\nu\to 0 }$ along the lower branch of the instability cone.    
%Let us assume that solutions exist for  $\Rn\to \infty$.  For the potential flow it is $\mu=\eta^2$. There is thus no other free parameter  describing the system than the inner radius $\eta$. It is then  suggestive that  the drift rate is also
%determined by a function of the inner radius $\eta$. The simplest possibility is the linear relation $\dot \phi/\Om_{\rm in} = \eta$ which is indeed confirmed by the numerical simulations. 
 On the other hand, Fig. \ref{fig2} (right) presents the values of (\ref{drift}) as
function of $\mu$ for three models with a wide, a medium and a narrow gap. The lines demonstrate a linear relation of the azimuthal drift and
the shear $\mu$, i.e.  
%They always start with the potential flow $\mu=\eta^2$. 
$\dot\phi/\Om_{\rm in}=\alpha(\mu-\eta^2)+\eta$ which leads to  $\dot\phi/\Om_{\rm in}=\eta$ for the potential flow with $\mu=\eta^2$. These drift values are indeed very  close to the diamonds positioned at the dotted line. 
The coefficient $ \alpha$ represents the slope between the normalized drift and the shear $\mu$.
%\beg
 %\frac{\dot\phi}{\Om_{\rm in}}\simeq \eta
% \label{eta} 
%\ende
To fix this slope  we  assume  $\dot\phi/\Om_{\rm in}=1$ for $\mu=1$   from the right panel of Fig. \ref{fig2} as valid for all $\eta$ 
and find 
\beg
 \frac{\dot\phi}{\Om_{\rm in}} = \frac{\mu+\eta}{1+\eta}.
 \label{eta2} 
\ende
For the container with $\eta=0.52$  a slope of $\alpha=0.66$ between drift and shear is   expected. With respect to the rotation
of the {\em outer} cylinder  (\ref{eta2}) leads to 
\beg
 \frac{\dot\phi}{\Om_{\rm out}}= \frac{1}{\mu}\ \frac{\mu+\eta}{1+\eta},
 \label{eta3} 
\ende
so that  ${\dot\phi}/{\Om_{\rm out}}= 1/\eta$ for the potential flow. Hence, for very thin gaps with $\eta\to 1$ this yields  ${\dot\phi}={\Om_{\rm out}}$
hence instability pattern and outer cylinder are exactly corotating in this limit. For containers with finite gaps    the pattern migrates slightly faster than the outer cylinder rotates. The differences to the rotation of the outer cylinder are a proxy of the gap width. Formally, for $\mu\gg \eta$ the relation $\dot\phi\simeq \Omout/(1+\eta)$ results independent of $\mu$. 

\begin{figure}
\centering
\includegraphics[width=0.85\textwidth]{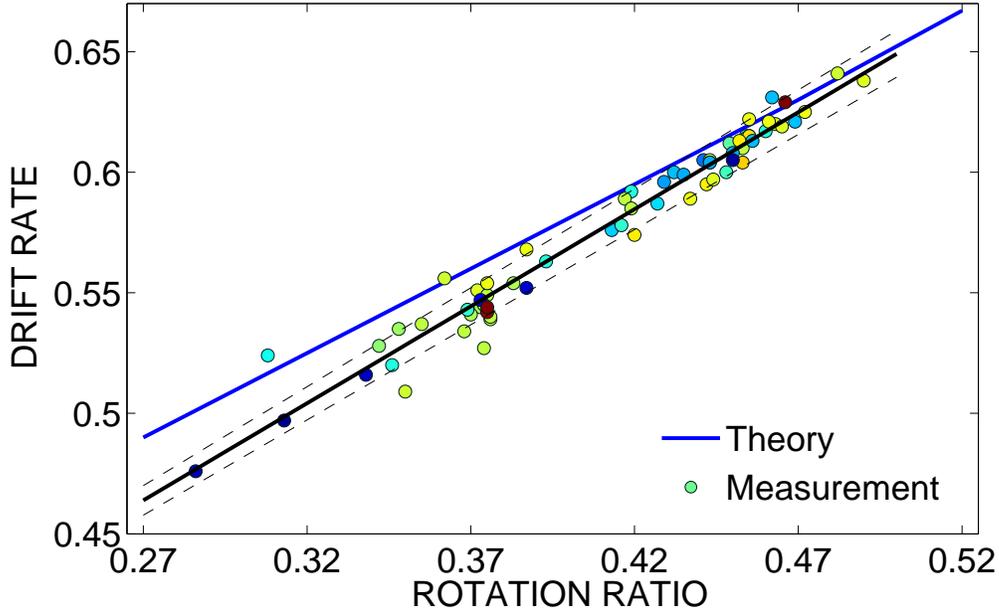}
\caption{Azimuthal drift rates (\ref{drift}) in comparison with the experimental results for  
$\Rn=150\pm 30$ (blue dots), $\Rn=210\pm 30$ (light blue dots), $\Rn=270\pm 30$ (green dots), $\Rn=330\pm 30$ (yellow dots),  $\Rn=390\pm 30$ (brown dots). The solid blue line gives the linear relation (\ref{eta2}).  The slope of the dark solid  line 
         representing the measurements is 0.77; between the two dashed lines $95\%$ of all measurements are located.  The slope of the blue line  is the same for all $\Rn$ intervals. The 
         drift always exceeds the rotation ratio $\mu$ so that the instability patterns  overturn the outer cylinder. $m=1$.}
\label{driftexp}
\end{figure}

Empirical results from  measurements with the SRI-container 
described in Section \ref{experiment} can be used to measure  the slope $\alpha$.
Figure \ref{driftexp} compares the  heuristical formulation  (\ref{eta2}) with  the empirical data. The  drift rate linearly runs with the shear parameter $\mu$. The more rigid the rotation
the closer are the  values of the observed  pattern drift and the calculated  results.  Differences between the numerical results and the
measurements mainly appear for the steepest rotation laws. The resulting observed  slope is $\alpha\simeq 0.77$. Indeed, Fig. \ref{fig2} (left) shows the drift of the potential flow with
$\mu=\eta^2$ being slightly smaller than the approximation (\ref{eta2}) predicts. It is however, not possible, to explain the deviations of the theoretical   and the
observed slope  only with  small inaccuracies  of the applied assumptions as they appear in Fig. \ref{fig2}. The relation (\ref{eta2}) concerns the horizontal lines in this  plot  for  slow-rotation but the values for the fast-rotation lines do hardly differ. The drift rates are uniform in the entire instability cones of Fig. \ref{fig1}.  For given rotation ratio $\mu$  the measurements shown in Fig. \ref{driftexp} fill the whole unstable interval.

In summary, the measured drift values are 
in  good agreement with those derived from the linear stability analysis.  Nonlinear simulations for models with endplates suggest that the deviations are produced by  the two endplates which have been ignored in formulating the relation (\ref{eta2}) but this unexpected result  is beyond the scope of this paper.

%%%%%%%%%%%%%%%%%%%%%%%%%%%%%%%%
\section{Flat rotation laws}
\begin{figure}
\centering
\includegraphics[width=0.85\textwidth]{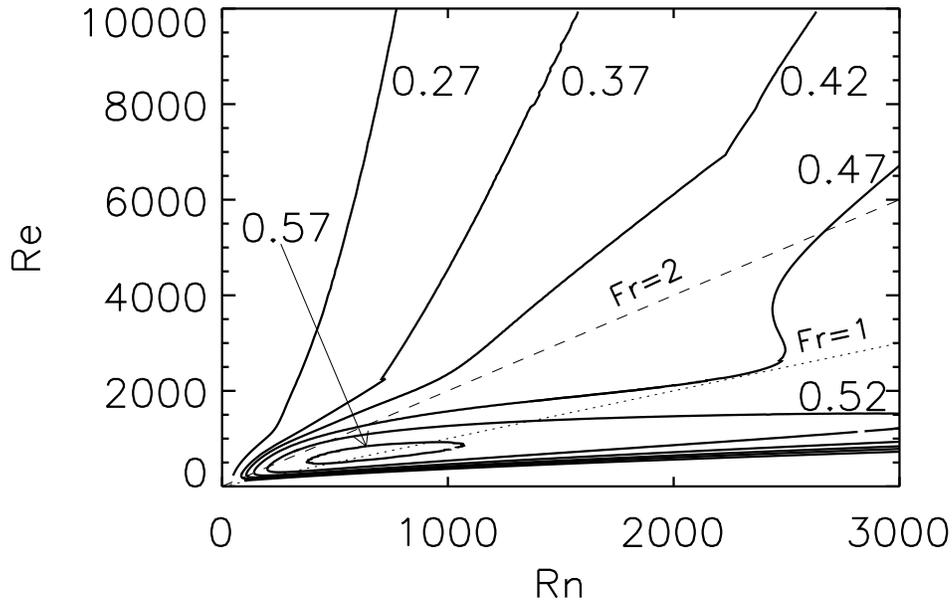} 
\caption{Stability maps of the $m=1$ mode in the  ($\Rn/\Rey$) plane. The dotted (dashed) lines mark $\Fr=1$ ($\Fr=2$). The curves are marked with their shear value $\mu$. The sign of the curvature of the 
         lines for fast rotation depends on $\mu$. The diffusive SRI disappears for $\mu_{\rm max}\simeq 0.57$ at $\Rey=\Rn=703$. $\eta=0.52$.}
\label{fig3}
\end{figure}

%%%%%%%%%%%%%%%%%%%%%%%%%%%%%%
%In their inviscid theory for unbounded flows Le Dizes \& Riedinger  (2010) find  flows with flat rotation laws  ($\mu>0.4 $)  as stable.
Figure \ref{fig3}  shows the   lines of marginal instability  for   rotation laws beyond the Rayleigh line,   $\mu=0.27$. The curves are  labeled with the shear parameter
$\mu$ of the differential rotation. For $\eta=0.52$  the quasi-Keplerian flow is represented by  $\mu=0.37$ and the quasi-uniform  flow $ U_\phi\simeq$~const  by $\mu=0.52$.  

We find all rotation laws for $\mu\leq \mu_{\rm max}$ with $\mu_{\rm max}=0.57$ to be unstable for moderate  Reynolds numbers in the  ($\Rn/\Rey$) plane.  The instability domain, however, continuously decreases for increasing $\mu$.
%, the potential flow possesses the largest domain of instability . 
For fixed stratification always two  Reynolds
numbers exist between them the flow is unstable. In particular, the flow is  stabilized by too fast rotation.  This rotational
stabilization is much stronger for the flat rotation laws than it is for the steeper profile   of the potential
flow. The transition from strong to weak stabilization lies somewhere between $\mu=0.42$ and $\mu=0.47$ where the curvature of the lines of neutral
instability changes its sign. For $\mu>0.52$ these lines encircle closed domains  which always contain the line  $\Fr=1$.  The  rotation profiles flatter than the rotation law of the quasi-uniform flow (where $\mu=\eta$) always possess even  an upper \BV number stabilizing the flow.

One finds two different forms of  the lines of neutral stability for  fixed 
$\mu$. They are closed if the rotation law is flat and they are open for steep  rotation laws.  
%The closed  domains with lower Reynolds numbers possess for the same value of $\mu$ open domains of instability but with much higher
%Reynolds numbers  which seem to be unbounded for $\Rey\to \infty$. The two domains are separated by a region of stability where the
%condition $\Fr=1$ does {\em not} lead to any linear instability of the mode  $m=1$. The phenomenon that certain moderate Reynolds numbers
%do not lead to an excitation of instability (while for lower $\Rey$ and for higher $\Rey$ they do) only occurs for the flat rotation 
%laws (Fig. \ref{fig3}). 
At $\mu_{\rm max}$  even  the closed instability domain disappears. For $\mu>\mu_{\rm max}$ the SRI can thus only be excited for
 Reynolds numbers larger than (say) $\mathcal{O}(10^4)$. As an example, one finds  $\Rey\simeq 60,000$ for instability of the rotation law with  $\mu=0.6$. 

%Note that $\mu_{\rm max}=0.57$ slightly exceeds the value $\mu_{\rm max}\simeq \eta=0.52$. Later
%calculations  and data by  \cite{IS16} showed that the maximum $\mu$ exceeds the value $\eta$, and the excess is limited here   to $\simeq 12\%$.  

\begin{figure}
\centering
\includegraphics[width=0.75\textwidth]{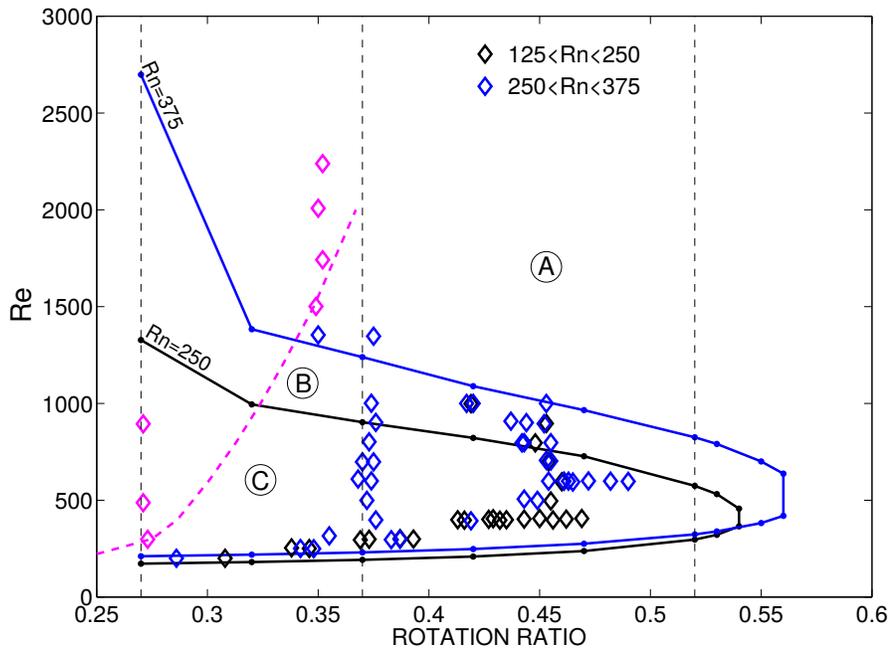}
\caption{Calculated lines of neutral stability  for the $m=1$ mode in the ($\mu/\Rey$) plane for  $\Rn=250$ (black) and   $\Rn=375$ (blue).  
         The vertical lines stand for $\mu=0.27$ (potential flow), for $\mu=0.37$ (quasi-Keplerian flow) and  $\mu=0.52$ (quasi-uniform  flow). 
         The empirical results of the Cottbus SRI experiment  are represented by  diamonds as the symbol of instability. The pink 
         line corresponds to the domain of strong endplate effects described in Fig. \ref{shear_inst}, also the measurements in this domain are marked by pink 
         symbols. $\eta=0.52$.}
\label{fig44a}
\end{figure}

\begin{figure}
\centering
\includegraphics[width=0.75\textwidth]{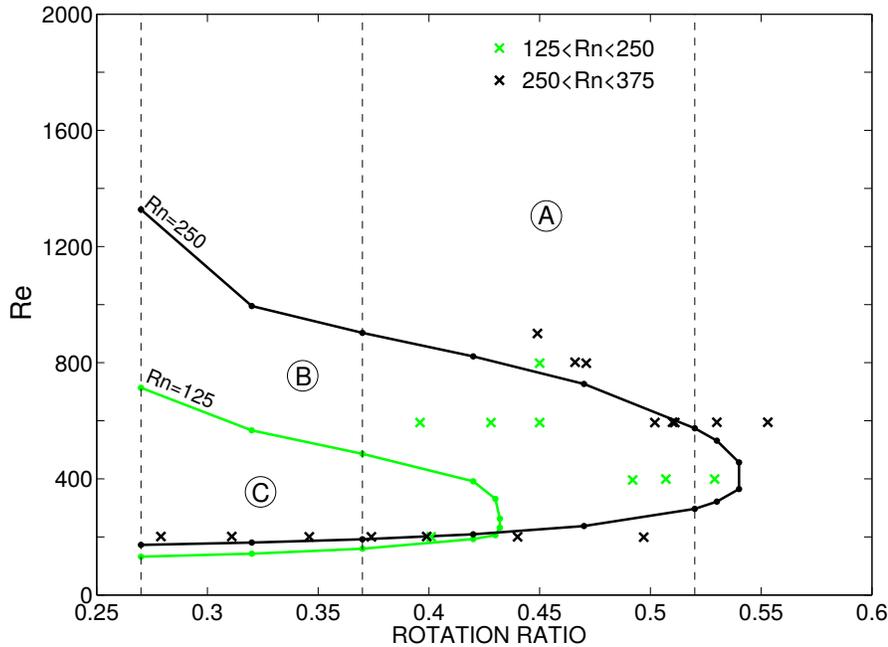}
\caption{Same as in Fig. \ref{fig44a} but for the experiments leading to  stable 
flows (crosses). Green line: neutral stability for  $\Rn=125$, black line: 
neutral stability for  $\Rn=250$.}
\label{fig44b}
\end{figure}

If the numerical results are mapped  in the ($\mu/\Rey$) plane   then the stability lines   for $\Rn=$~const have the
characteristic form shown in Figs. \ref{fig44a} and \ref{fig44b}  which have already  been observed  by \cite{IS16}. Different symbols  represent the empirical  
results 
%summarized  in Table \ref{Table} of the Appendix 
obtained in the experiment described in Section \ref{experiment}. The  diamonds in Fig. \ref{fig44a} used  for unstable flows 
demonstrate the existence of SRI for a wide range of shear values, Reynolds numbers and \BV numbers.  For 
given $ \mu$ and $\Rn$  a lower Reynolds number and an upper Reynolds number exist. Theoretically, all flows with $\Rn<250$ (black rhombs) 
should only be unstable in region C which is indeed the case. The blue rhombs  for flows with $250<\Rn<375$ should not appear in the region A  
since the rotation is there  too fast for the excitation of SRI.  
Probably because of endplate effects, the upper  Reynolds number limit is not well-defined. For fast rotation the size
of the pattern cells in axial direction would expand beyond the container  height so that  the system behaves no longer according to  the linear
 equations of the axially unbounded system. On the other hand, the measurements  precisely comply with  the numerically found  neutral line for 
slow rotation because of   its rather large wave numbers.

The crosses in Fig. \ref{fig44b} mark experiments without indication of SRI. They  define the domains of stability. Because of their definitions no black 
crosses should appear  in region  B and no crosses at all should appear in region C what is indeed the case.
The curves of constant $\Rn$ are always crossing each other. Note that there is a turning point in the upper parts of the curves. The curves are concave
for steep rotation laws and they are convex for flat rotation laws. Close to the Rayleigh line the stabilizing action of fast rotation is reduced.

The unstable rotation  laws  in the experiments fulfill $\mu\lsim 0.49$. The island solutions for $\mu \gsim 0.50 $ with their  low growth rates 
 have not been studied experimentally. The temperature differences of the used SRI experiment do not yet produce the 
needed high values of $Rn\simeq 700$.

%The formal reason is the change of the bending of the neutral lines with $\mu=$~const, shown in the top corner (left of $\Fr=2$) of Fig. \ref{fig3} which reflects the transition of open lines for steep rotation laws to closed lines for flat rotation laws. 

%%%%%%%%%%%%%%%%%%%%%%%%%%%%%%%%%%%%%%%%%%%%%%%%%%%%
\section{Cell size and azimuthal drift}
\begin{figure}
\centering
\hbox{\includegraphics[width=0.50\textwidth]{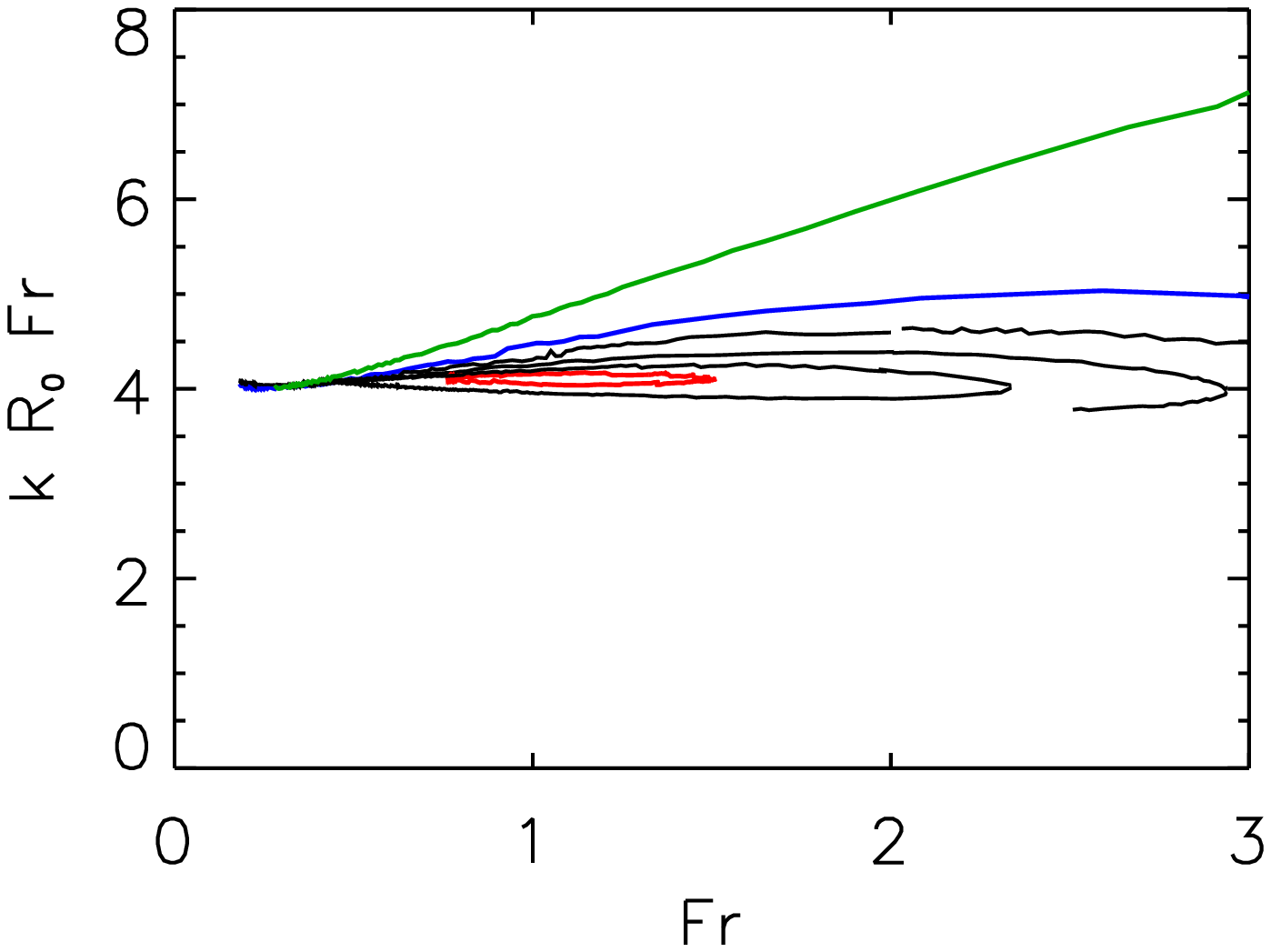}
      \includegraphics[width=0.50\textwidth]{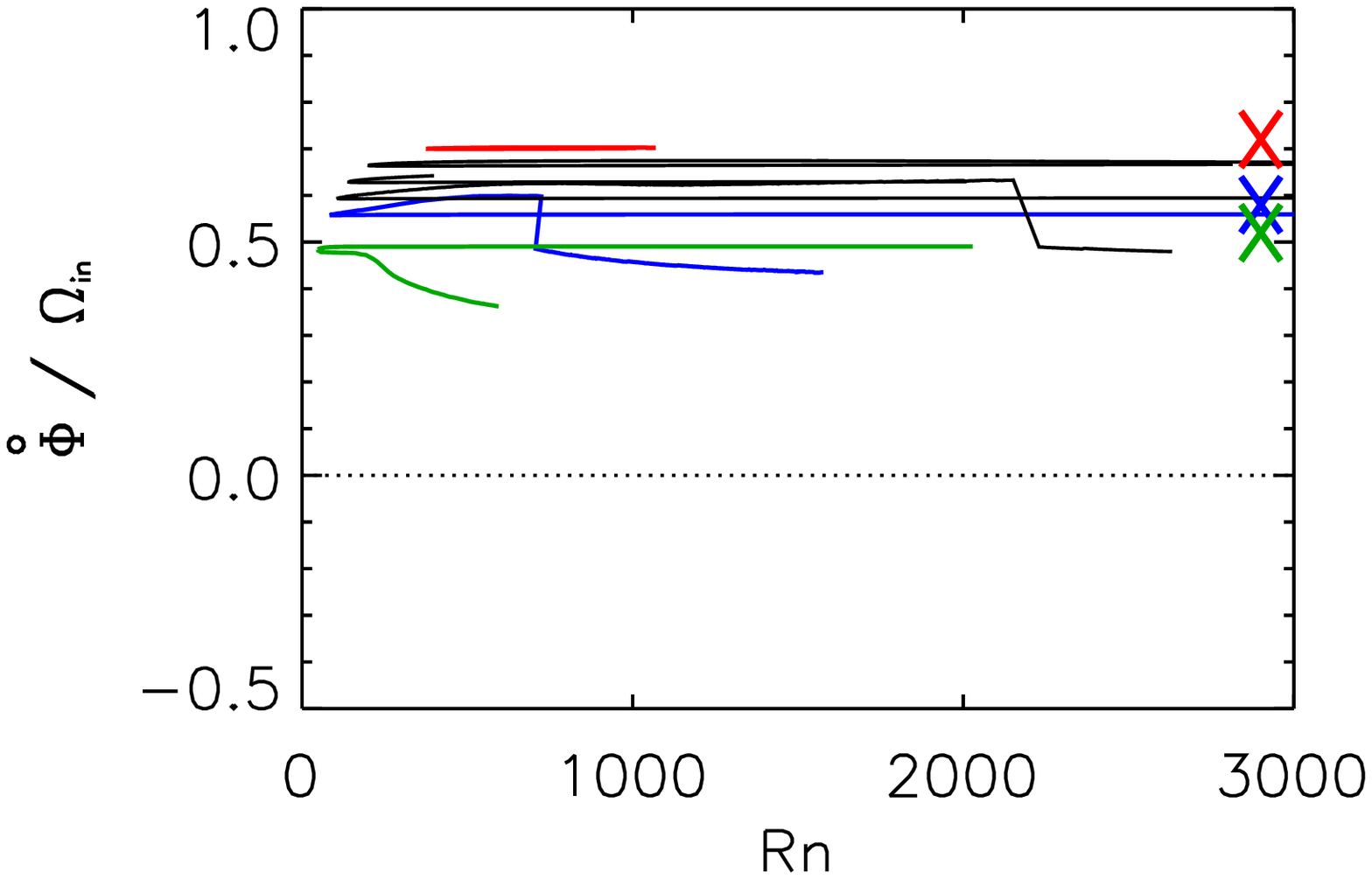}
     }
\caption{Left: Normalized axial wave numbers $k \Fr$ along the lines of marginal stability for  $\mu$ between 0.27 and 0.57 as a function of the Froude 
         number $\Fr$.   Right: Drift rates  (\ref{drift}).   The crosses symbolize the numerical values provided by Eq. (\ref{eta2}). Potential flow (green line), 
         quasi-Keplerian flow (blue line),  $\mu=0.57$ (red line). $m=1$, $\eta=0.52$. }
\label{fig4}
\end{figure}

%%%%%%%%%%%%%%%%%%%%%%%%%%%%%%%%%%%%%%%%%%%%%%%%%%%%
The axial size of a cell in units of  the gap width $D$ is
\beg
 \frac{\delta z}{D}\simeq \frac{\pi}{k}\sqrt{\frac{R_{\rm in}}{D}}
 \label{k}
\ende
so that for $D\simeq R_{\rm in}$ one finds $\delta z/D\simeq \pi/k$ with $k$ as the   wave numbers normalized with the radius $R_0$.  For large $k$  the cells are  oblate in axial direction. For $k\simeq \pi$ they are   nearly circular. Figure \ref{fig4} (left) gives the wave numbers along the lines of marginal stability. The green line compresses all  wave numbers for the potential flow given in Fig. \ref{fig1b} for $m=1$ in a surprisingly simple manner. $\Fr>0.5$ (fast rotation)   leads to the fixed wave
number $k\simeq 2$  while  smaller Froude numbers (slow rotation) lead to $k \Fr\simeq 4$.  For  $k\simeq 2$ the relation (\ref{k}) 
reads $\delta z/D \simeq 1.3$ leading to an axial cell size normalized with the size of the container,  $\delta z/H\simeq 1.3/\Gamma$, which for $\Gamma=10$ is certainly small enough even for experiments with  
fast rotation. 
%(but see the results in Fig. \ref{shear_inst}). 
For slow rotation the wave lengths are much shorter.

For the flows with weaker shear    we  find the wave numbers  depending 
on  $\Fr$ in the rather compact relation  $k  \Fr\simeq 4$ for all $\mu$, which with (\ref{k}) yields 
\beg
 \frac{\delta z}{D}\simeq \frac{\pi}{4}\ \frac{\Om_{\rm in}}{N}
 \label{deltaz} 
\ende
\cite{MM01}. In this formulation the action of the  Taylor-Proudman theorem is easy  to realize since for fast rotation   the
axial  cell sizes are larger than for slow rotation. The positive \BV frequency $N$ acts
 opposite: the weaker the stratification the more oblong are the cells. The wavelengths are short for slow rotation and large stratification. 
%In the experiments of \cite{IS16} the relation (\ref{deltaz}) has been probed with the result $\delta z/D\simeq \beta\cdot \Fr$ 
%with $\beta=0.86$ close to  $\beta\simeq \pi/4= 0.78$ taken from 
Equation (\ref{deltaz}) can be read as 
\beg
 \Fr= 1.28\ \Gamma\ \frac{\delta z}{H}
 \label{deltazz} 
\ende
relating the aspect ratio $\Gamma$ of a finite-height container to the  maximally reasonable $\Fr$. Provided the minimal number 
of cells in the container is (only) two then for $\Gamma=10$, the fastest possible rotation  follows from 
$\Fr\simeq 6.3$. With four cells required, $\Fr\simeq 3.1$ gives the upper limit. For the fast-rotation branch of the potential flow   
no such limit  exists. 

Again the drift rates of the patterns of marginal stability do not depend on the location in the ($\Rn/\Rey$) plane. The weak dependence on
the rotation law in Fig. \ref{fig4} (right panel)  approaches  the linear run with $\mu$ in accordance to (\ref{eta2}). The simplicity of
these results  is  amazing. Along the lines of marginal stability the drift values (\ref{drift}) do not vary and the
wave numbers are strictly anticorrelated to $\Fr$ with one and the same numerical factor with exception of the potentail flow.

%%%%%%%%%%%%%%%%%%%%%%%%%%%%%%%%%%%%%%%%%%
\section{Growth rates}\label{growth}
\begin{figure}
\centering
 \includegraphics[width=0.85\textwidth]{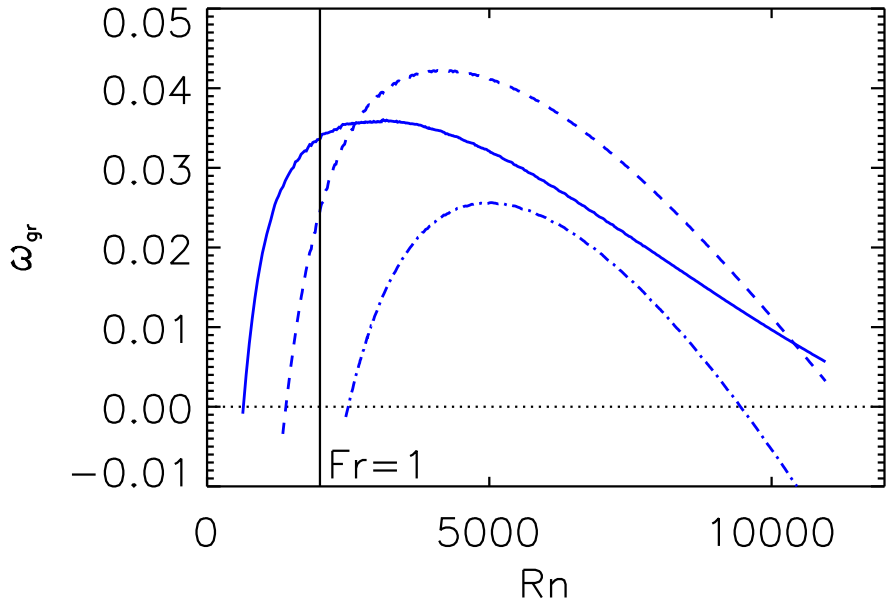}
 \caption{Growth rate  $\omega_{\rm gr}$ in units of $\Om_{\rm in}$   for 
 quasi-Keplerian flow with supercritical \BV~number and  fixed Reynolds number.   $m=1$ (solid), $m=2$ (dashed), 
          $m=3$ (dot-dashed). The vertical line marks $\Fr=1$. $\Rey=2000$, $\mu=0.37$.}
\label{fig8}
\end{figure}

%%%%%%%%%%%%%%%%%%%%%%%%%%%%%%%%%%%%%%%%%%
We briefly comment the growth rate $\omega_{\rm gr}$ of the instability which   is the imaginary part of the eigenfrequency  normalized with $\Om_{\rm in}$).  The   growth time   $\tau_{\rm gr}$ in units of 
the rotation time is $\tau_{\rm gr}/\tau_{\rm rot}=1/(2 \pi \omega_{\rm gr})$. 
To find the characteristic growth ratesof the instability of the quasi-Keplerian flow the  frequencies are computed for a given Reynolds number with supercritical $\Rn$. 
The Reynolds number
is fixed to $\Rey=2000$  for $\mu=0.37$ (Fig. \ref{fig8}).  One finds small  values of order 0.01, i.e. the growth time is typically 6
turnovers of the {\em outer} cylinder. The  maximal growth rates of the $m=1$ mode appear for $\Fr\lsim 1$. This result, however, depends on the mode number $m$. The higher modes grow fastest for smaller $\Fr$.  It is thus  possible  that
for larger  $\Rn$ the growth rate of a higher mode exceeds that for a lower mode.  

%%%%%%%%%%%%%%%%%%%%%%%%%%%%%%%%%%%%%%%
\section{Summary and conclusions}
\begin{figure}
\centering
\includegraphics[width=0.65\textwidth]{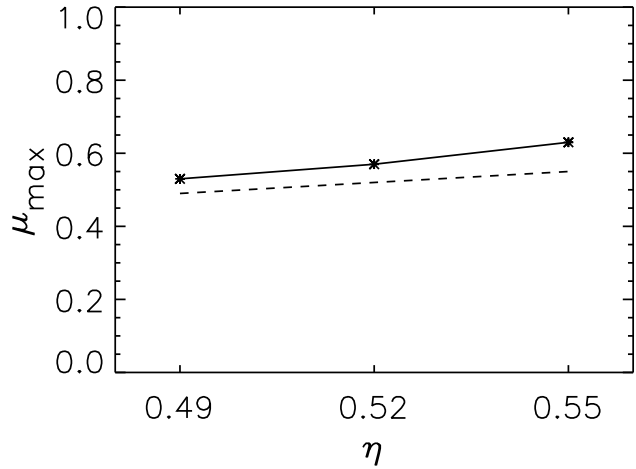}
\caption{The  theoretical maximal  values of $\mu$ for various gap widths $\eta$. The dashed line denotes the rotation law  with 
         $\mu=\eta$.  The calculated  $\mu_{\rm max}$  represent  rotation laws slightly flatter than $U_\phi=$~const. }
\label{fig45}
\end{figure}

%%%%%%%%%%%%%%%%%%%%%%%%%%%%%%%%%%%%%%%%
In view of the laboratory experiment described in Section \ref{experiment} the hydrodynamic equation system  (\ref{sysbo}) has been solved numerically for the
Taylor-Couette flow with the inner cylinder at $\eta=0.52$ for various rotation laws. The potential flow gives the simplest example  of the stratorotational instability. The lowest possible  Reynolds numbers are    only 135  for the potential 
flow ($\mu=0.27$) and  $\Rey=168$ for quasi-Keplerian flow ($\mu=0.37$).  There  is a well-defined maximal value of $\mu=\mu_{\rm max}$ at which the low-$\Rey$ regime suddenly stops. The maximum  $\mu_{\rm max}$ increases for narrower gaps.
For $\eta=0.52$ the value of $\mu_{\rm max}=0.57$ indicates that the characteristic rotation law is slightly flatter than the
quasi-uniform flow  with $U_\phi\simeq$~const. This finding remains true if the gap width is varied (Fig. \ref{fig45}).

Figure \ref{fig3} presents  the instability map for various   shear values $\mu$ of the  rotation laws  between the cylinders. It  shows the existence of a rotational stabilization of the SRI by fast rotation.  The instability (for fixed \BV number $\Rn$) only exists between a low and a large Reynolds number.
The  rotational stabilization is stronger for 
lower  $\Rn$  than for higher  ones. The effect increases for
increasing $\mu$, i.e. for flatter rotation laws.   For large $\mu$ the stability map even shows closed  domains so that for these rotation laws not only a maximum Reynolds number 
exists but also a maximum  \BV number. 
Such instability islands only  possesses  small  Reynolds numbers with $\mathcal{O}(10^3)$. We also know that for very large $\Rn $ and flat rotation laws 
 second instability areas occur  in the map.  For $\mu>\mu_{\rm max}$ this  instability type  requires Reynolds
numbers exceeding  $\mathcal{O}(10^5)$ which, however,  are not yet relevant for the present-day experimental possibilities.

Figures \ref{fig44a} and \ref{fig44b} show  the theoretical instability domains for all $\mu<\mu_{\rm max}$ for three \BV numbers $\Rn$.  There is indeed always a lower $\Rey$ and an upper $\Rey$ limiting the instability area. The dependence of the lower limiting $\Rey$ on 
$\Rn$ is very weak while it is much stronger for the upper limits. The experimental data  added to the plots match these intervals with only a few exceptions 
at the fast-rotation branches. The low-rotation limit is almost perfectly confirmed by the measurements. The definition of the upper limit for fast rotation 
is less precise. There are, in particular, unstable flows with Kepler-like rotation laws for Reynolds numbers which should already lead to stability based on 
the numerical results.  The wave numbers, however, at the upper line of neutral stability are so small that endplate effects by the apparatus 
(with aspect ratio $\Gamma=10$) must be expected. Note that in Fig. \ref{fig44b} all crosses correctly indicate the expected stability for the flatter 
rotation laws. 

Both azimuthal drifts and  wave numbers of the nonaxisymmetric modes fulfill simple rules. The patterns always drift in the positive
$\phi$-direction --  slightly faster than the outer cylinder rotates. Only for very narrow gaps  ($\rin\to 1$) the instability pattern 
would precisely corotate with the outer cylinder. The pattern speed does neither depend on the Reynolds numbers nor on the azimuthal mode 
number $m$. The drift data from the Cottbus SRI experiment indeed demonstrate  the rotation of the instability pattern as slightly 
faster than the rotation of the outer cylinder. The predicted linear relation between drift $\dot \phi$ and rotation ratio $\mu$ is 
also confirmed by the experiments. The theory  describes the observed azimuthal drift  phenomenon of the nonaxisymmetric perturbation patterns  
with acceptable  accuracy (see Fig. \ref{driftexp}).

One also finds a general rule for the wave numbers. Figure \ref{fig4}  provides  $R_0 k$ for the modes with $m=1$ in 
the regions of instability as basically behaving as $k\propto 1/\Fr$. The cells are thus nearly circular in the ($R/z$) plane
for $\Fr\simeq 1$. They are prolate for fast rotation with $\Fr>1$ and they are oblate for slow rotation with $\Fr<1$.  As an exception the potential flow with $\Fr>1$ shows uniform values of the wave number independent of $\Fr$. Nevertheless, to describe the shape of the instability cells is the true role of the Froude number in the theory of stratified Taylor-Couette flows.
%As an 
%exception, the cells for the potential flow instability are always nearly circular in the ($R/z$) plane independent of the actual Froude number.

%\begin{acknowledgments}
%\ack
%M.G. acknowledges  support from the Helmholtz alliance {\em LIMTECH}.
%T.S., U.H. and C.E. thank the Deutsche Forschungsgemeinschaft (DFG) for the grants  HA 2932/7-1 and EG100/18-1.
%U.H. thanks DFG for support via the MS-GWaves research unit HA 2932/8-1.
%\end{acknowledgments}
\bigskip

%The behavior of  the drift rates along the line $\Rey=2000$ is also   characteristic. The drift rates (\ref{drift})  do  not depend on  $m$ nor on
%the value of $\Rn$. The drift rates are  equal for the modes with various azimuthal mode numbers (Fig. \ref{fig8},
%right).

%\appendix{\bf Appendix}
%\input{table.tex}
%\newpage

\end{document}